\newcommand{\be}{\begin{equation}}
\newcommand{\ee}{\end{equation}}
\newcommand{\bea}{\begin{eqnarray}}
\newcommand{\eea}{\end{eqnarray}}

\documentstyle[preprint,eqsecnum,aps]{revtex}
\begin{document}
\draft
\preprint{Alberta-Thy-3-94}
\title{THE BOULWARE STATE AND THE GENERALISED SECOND LAW OF
  THERMODYNAMICS}
\author{Warren G. Anderson}
\address{Theoretical Physics Institute, University of  Alberta,\\
   Edmonton, Alberta, Canada T6G 2J1}
\maketitle
%%%%%%%%%%%%%%%%%%%%%%%%%%%%%%%%%%%%%%%%%%%%%%%%
%
%ABSTRACT
%
%%%%%%%%%%%%%%%%%%%%%%%%%%%%%%%%%%%%%%%%%%%%%%%%

\begin{abstract}
We show that the appropriate vacuum state for the interior of a box with
reflecting walls being lowered adiabatically into a Schwarzschild black hole is
the Boulware state.  This is concordant with the results of Unruh and Wald, who
used a different approach to obtain the stress-energy inside the box. Some
comments about an entropy bound for ordinary matter, as first conjectured by
Bekenstein, are presented.
\end{abstract}
\clearpage

\narrowtext
%%%%%%%%%%%%%%%%%%%%%%%%%%%%%%%%%%%%%%%%%%%%%%%%
%
%INTRODUCTION
%
%%%%%%%%%%%%%%%%%%%%%%%%%%%%%%%%%%%%%%%%%%%%%%%%
\section{INTRODUCTION}  \label{sec:intro}
One of the most remarkable developments in gravitational theory in the last
century has been the discovery that fields quantised on a black hole background
exhibit thermodynamical properties\cite{Haw:74}. This discovery was presaged by
the work of Bardeen, Carter and Hawking\cite{BCH:73} in which they pointed out
an analogy between laws governing certain properties of black holes and the
laws of ordinary thermodynamics. In particular, the analogue of the second law
of thermodynamics is Hawking's theorem that the surface area of a black hole is
nondecreasing \cite{Haw:71}, i.e.
\be
    \frac{dA_{BH}}{d\tau} \ge 0.  \label{eq:1}
\ee

It was based on this analogy between (\ref{eq:1}) and the second law of
thermodynamics that Bekenstein\cite{Bek:73} conjectured a generalised second
law of thermodynamics (GSL): {\em The sum of the black hole entropy and the
ordinary entropy in the black hole exterior never decreases}. More precisely,
the GSL states that for any physical process
\be
     \delta S_{matter} + \frac{1}{4} \delta A_{BH} \ge 0, \label{eq:2}
\ee
(units $\hbar$=c=G=k=1),
where $S_{matter}$ is the entropy of the matter outside the black hole. In
(\ref{eq:2}), $\frac{1}{4} A_{BH}$, one quarter of the black hole's surface
area, plays the role of the entropy of the black hole. This correspondence
between the surface area and entropy of a black hole has become firmly
established in the context of black hole thermodynamics, beginning with the
Hawking's discovery of the thermal radiation emitted by a black
hole\cite{Haw:74}.

Bekenstein\cite{Bek:81} further argued that an entropy bound on matter was
required in order for the GSL to hold. His argument relied on the following
{\em Gedankenexperimente}. Let us imagine that a box of linear dimension R with
reflecting walls is filled with ordinary matter of energy $E_{\infty}$ and
entropy $S$ at a very large proper distance from a black hole. The box is then
slowly (adiabatically) lowered toward the black hole of mass $M$.  When the box
is opened and the matter released into the black hole, the energy of the matter
will have been reduced by the redshift factor $\chi = (1-2M/r)^{1/2}$ so that
the black hole's energy is increased by
\be
     E = \chi E_{\infty}. \label{eq:3}
\ee
Since we can lower the box to approximately the distance R (the dimension of
the box) from the event horizon before releasing the energy into the black
hole, we can provide the black hole with as little as $E=(1-2M/(R+2M))^{1/2}
E_{\infty}$ energy. However, as Bekenstein demonstrated, this will lead to a
change in the black hole entropy of
\be
     \delta S_{BH} = \frac{1}{4} \delta A_{BH} = 8 \pi M E. \label{eq:4}
\ee
After the box is emptied, it can be slowly pulled back out to infinity. But
observe that, if $R<S/(2\pi E_{\infty})$, we will have $\delta S_{BH} < S$ and
the GSL will be violated. Therefore, Bekenstein concluded there was a bound on
the entropy of matter with energy $E$ that could be placed in a box of
dimension $R$,
\be
      S/E\le 2 \pi R.     \label{eq:5}
\ee

Unruh and Wald\cite{U&W:82} (UW) have pointed out, however, that Bekenstein
failed to consider black hole quantum effects in his analysis. In particular,
they point out the effect of acceleration radiation on the box as it is being
lowered. Since, in the reference frame of the almost stationary (hence
accelerated) box, the black hole is surrounded by a bath of thermal radiation,
there will be an upward pressure on the box. In fact, when this is taken into
account, Unruh and Wald demonstrate that the box will float when the energy
contained in the box, $E$, is exactly the same as the energy of the
acceleration radiation displaced by the box. In order to lower the box further,
one will have to do work against this buoyancy force. Unruh and Wald go on to
show that in order to minimize the entropy increase of the black hole, the box
must be opened at the floating point. They further show that the matter
released at this point will contribute at least enough energy to the black hole
to increase its entropy by an amount
\be
     \delta S_{BH} \ge S,     \label{eq:6}
\ee
where S is the entropy of the matter in the box. Thus, they conclude, the GSL
will hold independently of the validity of (\ref{eq:5}).

More recently, Li and Liu\cite{L&L:92} have stated that the belief of Unruh and
Wald that the Hawking radiation is thermal near the black hole is in error. In
support of this statement they use the approximate stress-energy tensor for a
massless scalar field surrounding a black hole found by Page\cite{Pag:82}. They
demonstrate that this stress-energy does not have the form of a perfect gas of
photons in thermal equilibrium. They go on to derive a new equation of state
for the Hawking radiation near the black hole. They find that the pressure of
the Hawking radiation near the black hole is not large enough to produce a
substantial buoyancy effect, and derive a bound on the entropy very similar to
(\ref{eq:5}) of Bekenstein.

Page's approximate stress-energy is for the Hartle-Hawking state associated
with a conformally coupled massless scalar field in a Schwarzschild background.
This is an adequate description of the state {\em outside} the box as seen by a
{\em freely falling} observer. However, as demonstrated in UW, because the box
is accelerating, it is subject to a bath of {\em acceleration} radiation.
Further, UW show that this acceleration can be thought of as affecting the
energy density {\em inside} the box.

How can the acceleration of the reflecting walls of the box affect the energy
inside? Let us first answer this question for a box of fixed proper length
accelerating in flat space. It is well established that, when quantum effects
are considered, a mirror experiencing nonuniform acceleration will radiate two
fluxes of energy proportional to the  change in acceleration, $dE \propto
da$\cite{F&D:76}. One of these fluxes will be in the direction of the change in
the acceleration of the mirror, and will have negative energy. The other, in
the opposite direction, will have positive energy.

We first consider the situation from the point of view of an inertial observer
watching the mirrored box accelerate from left to right. If the box increases
its acceleration, two fluxes of energy will enter the box. The flux from the
rear (left) wall will be negative and the flux from the front (right) wall will
be positive. However, these fluxes will not be equal. As the box accelerates,
it will undergo Lorentz contraction. The rear wall will therefore be forced to
accelerate, and change its acceleration, at a higher rate than the front wall,
and will thus emit a larger flux. As a result, the inertial observer sees a
negative energy density developing inside the box.

Now, let us consider the situation from the point of view from an observer
inside the box, accelerating with it. This observer does not notice a negative
energy density developing inside the box. Indeed, this observer, who started in
the empty Minkowski vacuum, still believes that the interior of the box is
(apart from himself) empty. Thus, with respect to what he sees as the vacuum
state, the exterior of the box is filled with a positive energy fluid. This
fluid is none other than the acceleration radiation described by UW. It should
be emphasised that the bath of acceleration radiation seen by the accelerating
observer is an artifact of this observer measuring energy with respect to the
vacuum of his non-inertial (accelerated) frame.  The true stress-energy for the
quantum fields and accelerated mirrors in flat space is properly described by
the inertial observer, who sees a negative energy density inside the box.

Let us now return to the case of the rigid box being lowered toward a black
hole. Both the top and bottom reflecting walls will undergo a change of
acceleration. The positive energy flux from both mirrors will be toward the
horizon, the negative energy fluxes away from the horizon. Thus, as we lower
the box, positive energy will flow from the mirror at the top of the box into
the boxes interior, while at the same time, negative energy will flow from the
bottom mirror into the box. But for a box of fixed proper length, the change in
acceleration during lowering is larger at the bottom than at the top.
Therefore, the flux from the bottom mirror will be larger, and there is a net
negative energy flow into the box. The interior of a box which is initially
empty will consequently acquire a negative energy density through the lowering
process. This negative energy density is exactly what one would expect from the
Boulware state.

Using a 1+1 dimensional model, which we believe captures the essential features
of the problem, we will show that energy of the contents of the box is indeed
correctly measured with respect to the (negative) energy of the Boulware state.
Furthermore, it is easy to show that the measurement of the box's internal
stress-energy with respect to the Boulware state is in full agreement with UW.
We wish to stress that we obtain the Boulware vacuum energy from considering
only the acceleration of the reflecting walls of the box in a {\em flat}
background. This, we feel, is a remarkable result which may be exploited more
fully in the future.

%%%%%%%%%%%%%%%%%%%%%%%%%%%%%%%%%%%%%%%%%%%%%%%%
%
%1+1 black hole Vacua
%
%%%%%%%%%%%%%%%%%%%%%%%%%%%%%%%%%%%%%%%%%%%%%%%%
\section{1+1 DIMENSIONAL BLACK HOLE VACUUM STATES}
Let us begin by considering a 1+1 dimensional black hole with metric
\bea
     ds^2&=&-f(r)dt^2+\frac{dr^2}{f(r)}\nonumber\\
     f(r_0)&=&0,f'(r_0)=2\kappa \label{eq:7}
\eea
where $r_0$ is the horizon radius, $f'$ denotes $df/dr$, and $\kappa$ is a
constant. The redshift factor for metric (\ref{eq:7}) is $\chi=\sqrt{f}$ and
$dz:=dr/\sqrt{f}$ defines $z$ which measures the proper distance from the
horizon.

Let us introduce null coordinates $u$ and $v$ defined by
\bea
     dv &:=& dt+\frac{dr}{f(r)} = -n_a dx^a, \nonumber \\
     du &:=& dt-\frac{dr}{f(r)} = -l_a dx^a, \label{eq:7.1}
\eea
where indices $a,b,c,\ldots$ range over $0,1$.
The expectation value for the stress-energy tensor of a massless field on the
background (\ref{eq:7}) can be written in the form
\be
     T^{ab} = \frac{1}{2} T^c_c g^{ab} + E(l^al^b+n^an^b) + Fl^al^b,
\label{eq:7.2}
\ee
where the unspecified functions $T^c_c(r), E(r)$ and $F(r)$ correspond to
vacuum polarisation, an isotropic radiation field and a net outward flux
respectively.

For a massless scalar field, the function $T^c_c$ is given by the ``trace
anomaly'',
\be
     T^c_c=\frac{1}{24\pi} R = - \frac{1}{24\pi} f'',  \label{eq:7.3}
\ee
where $R$ is the curvature scalar for metric (\ref{eq:7}). One can obtain the
remaining components from the conservation law, $T^{ab}{_{|a}}=0$, where $|$
denotes covariant differentiation. In terms of $E$ and $F$ the conservation law
takes the form
\bea
     F'(r)&=&0, \nonumber \\
     E'(r)&=&-\frac{1}{4}f(r)\frac{d}{dr}T^c_c(r). \label{eq:7.4}
\eea
The specific vacuum state with respect to which the expectation value of the
stress-energy tensor is taken is given by the boundary conditions which are
imposed on (\ref{eq:7.4}).

We will be interested in two types of vacuum stress-energy here. The Boulware
state appears empty, apart from the vacuum polarisation represented by
(\ref{eq:7.3}), to stationary observers. This is expressed by the boundary
condition $T^{ab} \rightarrow 0$ as $r \rightarrow \infty$. This condition and
the conservation equations (\ref{eq:7.4}) imply
\bea
     E \equiv E_B &=& \frac{1}{48\pi}\left( \frac{1}{2} f f'' -\frac{1}{4} f'^2
\right), \nonumber\\
     F \equiv F_B &=& 0. \label{eq:7.5}
\eea
When (\ref{eq:7.3}) and (\ref{eq:7.5}) are substituted into (\ref{eq:7.2}) the
stress-energy takes the form of a stationary fluid with energy density and
pressure
\bea
     \rho_B &=& \frac{1}{24\pi}\left(f''-\frac{f'^2}{4f}\right),
\label{eq:7.6}\\
     P_B &=& -\frac{1}{24\pi}\frac{f'^2}{4f}, \label{eq:7.7}
\eea
respectively.

The Hartle-Hawking state is the one which is appropriate for an eternal black
hole inside a cavity with reflecting walls, in thermal equilibrium with its own
radiation. It appears empty (modulo vacuum polarisation) to free-falling
observers at the horizon. This corresponds to the boundary condition that the
stress-energy be regular on both the past and future event horizons. By
imposing this boundary condition on equations (\ref{eq:7.4}) we find that $E$
and $F$ take the form
\bea
     E \equiv E_{HH} &=& \frac{1}{48\pi} \left( \frac{1}{2} f f'' +
\kappa^2-\frac{1}{4}f'^2\right)
          \nonumber \\
     F \equiv F_{HH} &=& 0 \label{eq:7.8}
\eea
Thus, the expectation value of the stress-energy in the Hartle-Hawking state
also takes the form of a stationary fluid with energy density and pressure
\bea
     \rho_{HH} &=& \frac{1}{24\pi}\left(f''-\frac{4\kappa^2-f'^2}{4f}\right),
\label{eq:7.9}\\
     P_{HH} &=& -\frac{1}{24\pi}\frac{4\kappa^2-f'^2}{4f}, \label{eq:7.10}
\eea
respectively. Notice that as $r \rightarrow \infty$ we have $P_{HH} \approx
\rho_{HH} \approx \kappa^2/24\pi$. This is the thermodynamical equation of
state for black-body radiation at temperature
\be
     T \equiv T_{BH}=\kappa/2\pi. \label{eq:7.11}
\ee
$T_{BH}$ is taken to be the temperature of the black hole.

%%%%%%%%%%%%%%%%%%%%%%%%%%%%%%%%%%%%%%%%%%%%%%%%
%
%Acceleration radiation and the Boulware Vacuum
%
%%%%%%%%%%%%%%%%%%%%%%%%%%%%%%%%%%%%%%%%%%%%%%%%
\section{ACCELERATION RADIATION AND THE BOULWARE STATE}  \label{sec:BoulVac}
In UW, the energy flux from a accelerating mirror in 1+1 dimensions is
considered. They find, in accordance with Fulling and Davies(\cite{F&D:76}),
that $dE=-da/12\pi$ from which we obtain
\be
     \frac{dE}{dz}=-\frac{1}{12\pi} \frac{da}{dz}.   \label{eq:8}
\ee
For the spacetime (\ref{eq:7}), the magnitude of the four acceleration, $a$, at
a proper distance $z$ from the horizon is given by
\be
     a=\frac{1}{\chi} \frac{d\chi}{dz} = \frac{f'}{2\sqrt{f}}. \label{eq:9}
\ee
Thus, the magnitude of the energy flux from a mirror is
\be
     dE=-\frac{1}{24\pi}\left(f''-\frac{f'^2}{2f} \right)dz. \label{eq:10}
\ee

Now, let us consider a rigid box with reflecting walls being lowered
adiabatically toward the black hole. Let us assume the top and bottom walls are
rigidly separated by a proper length $\ell$ which is much less than the radius
of the black hole. At the surface labeled by $z_0$ in the interior of the box,
the energy due to the acceleration radiation will be the energy from the flux
from the top of the box, blueshifted to the appropriate value, plus the energy
from the flux from the bottom of the box, redshifted to the appropriate value,
\be
     dE_{net}(z_0) = \frac{1}{24\pi}\left\{
     \frac{\sqrt{f(z_T)}}{\sqrt{f(z_0)}}\left( f''-\frac{f'^2}{2f}
     \right)_T-\frac{\sqrt{f(z_B)}}{\sqrt{f(z_0)}}\left(f''-\frac{f'^2}{2f}
      \right)_B  \right\} dz,
          \label{eq:11}
\ee
where the subscripts $0$, $B$, and $T$ denote quantities at $z_0$, the bottom,
and the top of the box respectively. In obtaining (\ref{eq:11}) we have taken
advantage of the fact that since the proper length of the box, $\ell$, is
assumed constant, $z_B=z_T+\ell$, and therefore $dz_B=dz_T=dz$.

For small $\ell$,
\be
     \frac{dF}{dz} \simeq \frac{F(z+\ell)-F(z)}{\ell}, \label{eq:12}
\ee
 for any function $F(z)$. Therefore, we rewrite (\ref{eq:11}):
\be
     dE_{net}(z_0) \simeq \frac{\ell}{24 \pi \sqrt{f}}
     \left\{\frac{d}{dz}\left[\sqrt{f}\left(
     f''-\frac{f'^2}{2f}\right)\right]\right\}dz. \label{eq:13}
\ee
The total energy is the sum of the energies entering the box as it is lowered
from infinity to $z_0$,
\bea
     E_{net}(z_0) &\simeq& \frac{\ell}{24\pi}\int_{\infty}^{z_0}
     \frac{1}{\sqrt{f}}\left\{\frac{d}{dz}\left[\sqrt{f}\left(
     f''-\frac{f'^2}{2f}\right)\right]\right\}dz
          \nonumber \\
     &\simeq& \frac{\ell}{24\pi}\left(f''-\frac{f'^2}{4f}\right)_{z_0}
\label{eq:14}
\eea
But this is just the energy of the Boulware state (\ref{eq:7.6}). Thus,  the
energy of the matter content of the box is properly measured with respect to
the Boulware vacuum energy.

Alternatively, let us examine the forces involved in lowering the box toward
the black hole.
By taking the Boulware state to be the vacuum state inside the box, and the
Hartle-Hawking state to be the state outside the box, one can recover the
results of UW.  To prove this, let us consider the contribution of radiation
pressure on the top and bottom of the box to the force needed to lower it. The
pressure on each of the reflecting walls is the difference between the
Hartle-Hawking fluid on the outside of the box and the Boulware fluid on the
inside,
\be
     P_{net} = P_{HH}-P_{B} = \frac{\kappa^2}{24\pi f}. \label{eq:15}
\ee
Thus the net contribution to the force needed to lower the box is
\be
     F_{net} = \frac{\kappa^2}{24\pi}\left(\frac{1}{f}_T - \frac{1}{f}_B\right)
          =\frac{\pi}{6}(T_T^2-T_B^2),  \label{eq:16}
\ee
where $T_T=T_{BH}/\chi_T$ and $T_B=T_{BH}/\chi_B$ are the redshifted values of
the black hole temperature at the top and bottom of the box respectively. This
is precisely the expression obtained by UW for the contribution of the
acceleration radiation to the force needed to lower the box.

%%%%%%%%%%%%%%%%%%%%%%%%%%%%%%%%%%%%%%%%%%%%%%%%
%
%Conclusion
%
%%%%%%%%%%%%%%%%%%%%%%%%%%%%%%%%%%%%%%%%%%%%%%%%
\section{CONCLUSION}  \label{sec:Conclusion}
We have examined once more the long debated question of whether the validity of
the GSL implies an entropy bound of the type found by Bekenstein (\ref{eq:5}).
We considered the standard {\em Gedankenexperimente} of lowering a box
containing matter fields from infinity to a finite proper distance from a black
hole. We have shown that the effect of the acceleration radiation on the energy
density inside the box is exactly the same as obtained by considering the
vacuum state of the interior of the box to be the Boulware state, and concluded
that the acceleration of the box has induced a Boulware state inside it.

That the in-vacuum for the box's interior is the Boulware state is to be
expected on general grounds. The interior vacuum is initially the Boulware
state (the vacuum for a stationary observer at infinity) and is invariant under
the adiabatic (quasi-static) process of lowering. This is in contradiction with
Li and Liu\cite{L&L:92}, who, by ignoring the effect of acceleration radiation,
implicitly assume a Minkowski vacuum inside the box. We have, furthermore,
shown that using the Boulware state for the interior of the box leads to
complete agreement with he results of UW.  This is not surprising since we
started with a relation derived by UW, (\ref{eq:8}), in our derivation of the
Boulware energy density for the interior of the box.

What may be surprising is that the acceleration of an empty box in flat
spacetime can be used to obtain the energy density of the Boulware state.
However, this is not difficult to understand. As we mentioned, it is expected
that the state inside an adiabatically lowered box is the Boulware state.
However, a small enough box sees the black hole field as being essentially
homogeneous. By the equivalence principle, such a box is unable to determine
whether it is accelerating in a gravitational field or in Rindler
space\cite{C&S:82}. We have demonstrated how this allows us to obtain the
Boulware energy density simply by considering the energy balance between two
accelerating mirrors in 1+1 dimensional Rindler space. We are presently
investigating possibility of carrying out this procedure in 3+1 dimensions.

\acknowledgments
I am indebted to Werner Israel for providing the inspiration for, and much
insight into, this paper. I would also like to thank Pat Brady and Geoff
Hayward for there helpful comments.

\end{document}